\documentclass[twocolumn,showpacs,preprintnumbers,amsmath,amssymb]{revtex4}

\usepackage{fancyheadings}
\usepackage{amsmath}
\usepackage{amsfonts}
\usepackage{amssymb}
\usepackage{graphicx}
\begin{document}

\title{Models of Topology Change}
\author{Alfred D. Shapere$^1$, Frank Wilczek$^{2,3}$, and Zhaoxi Xiong$^3$\vspace*{.1in}}
\affiliation{$^1$Department of Physics and Astronomy, 
University of Kentucky, Lexington, Kentucky 40502 USA\\
$^2$Center for Theoretical Physics,
$^3$Department of Physics, Massachusetts Institute of Technology,
Cambridge, Massachusetts 02139 USA}

\vspace*{.3in} 
\begin{abstract}
We show how changes in unitarity-preserving boundary conditions allow continuous interpolation among the Hilbert spaces of quantum mechanics on topologically distinct manifolds.  We present several examples, including a computation of entanglement entropy production.  We discuss approximate realization of boundary conditions through appropriate interactions, thus suggesting a route to possible experimental realization.   We give a theoretical application to quantization of singular Hamiltonians, and give tangible form to the ``many worlds'' interpretation of wave functions.
\end{abstract}

\maketitle

\thispagestyle{fancy}

\bigskip

%%%%%%%%%%%%%

Quantum theory, starting with its name, has a reputation for introducing discreteness.  It can, however, also blur it.
Here we show how the behavior of quantum mechanical models based on given manifolds can be interpolated continuously, as the manifolds undergo topological changes.   

The application of quantum mechanics to general relativity suggests the possibility of violent fluctuations in geometry under extreme circumstances, particularly in virtual processes at ultra-short distances.  It has widely been speculated that changes in spacetime topology might occur as a result.    Unfortunately it appears very difficult to perform detailed or convincing dynamical calculations for quantum gravity, even in the possibly consistent form of string theory, or to adduce relevant phenomena.  (The joining and splitting of strings themselves is a form of topology change, whose description involves the special, complicated apparatus of vertex functions and string field theory.)   
But as we will describe, in the relatively mundane context of quantum mechanics, topology change amounts to no more than a smooth change of boundary conditions in  Hilbert space, or, equivalently, a change of parameters in the Hamiltonian.  The concrete framework we present may suggest a less singular, more inherently quantum mechanical approach to topology change, and perhaps shed light on the emergence of spacetime itself.  It is not implausible that systems of the sort we consider could be realized experimentally. 

The idea of changing topology by smoothly varying boundary conditions in Hilbert space was proposed in Ref.~\cite{bala} and further developed in Ref.~\cite{Asorey}.  In \cite{bala} it was pointed out that topologically distinct boundary conditions consistent with unitarity correspond to different self-adjoint extensions \cite{vonNeumann}\cite{ReedSimon} of the Hamiltonian, between which there exists (at least in simple cases) a continuous interpolation through the space of possible unitary boundary conditions.

Here we take a concrete approach to topology change, in which boundary conditions are implemented in Hilbert space by projection operators, and develop simple examples in detail.    For the most basic example, the fission of one interval into two, we follow the behavior of the eigenfunctions under adiabatic interpolation between the two topologies, and compute the entanglement entropy produced by this process.  We also consider an application to a singular Hamiltonian system, which admits multiple topologically inequivalent quantizations.  Finally, we discuss the implementation of boundary conditions via direct modification of the Hamiltonian, and close with some comments regarding possible further applications. 

{\it Two-Point Unitary Boundary Conditions\/}:  To begin, consider the quantum mechanics of a particle supported on a space parameterized as two rays $-\infty < x \leq a$, $-\infty < x \leq  b $.   We assume that the kinetic energy density is given, away from the endpoints, by $| {\partial_x \psi} |^2$; a smooth potential is also allowed.  The Schr\"odinger equation gives us a probability current $j = i (\psi^* {\partial_x \psi} - {\partial_x \psi^*\,} \psi)$.   To insure that no probability flows out at $a, b$, we need a linear boundary condition satisfying $j(a) = j(b) = 0$.   Any two Robin boundary conditions
\begin{eqnarray}\label{robin}
0 ~&=&~ \alpha \psi (a) + \beta \frac{\partial \psi}{\partial x} (a) \nonumber \\
0 ~&=&~ \gamma \psi (b) + \delta \frac{\partial \psi}{\partial x} (b)
\end{eqnarray}
with fixed real $\alpha, \beta, \gamma, \delta$ do the job.   Conventional Dirichlet ($\beta, \delta = 0$) or Neumann ($\alpha, \gamma =0$) boundary conditions are of course included as special cases.   The conditions (\ref{robin}) also are sufficient to give a good eigenvalue problem for the Hamiltonian, since they are appropriate to terminate a second-order differential equation.   If we impose boundary conditions of the type (\ref{robin}) separately at both $a$ and $b$, we will describe the dynamics of a particle that lives in two independent spaces, with a fixed (by initial conditions) probability to be found on either one.   At another extreme, we can also obtain unitary dynamics and a good eigenvalue problem by imposing two ``identification'' conditions
\begin{eqnarray}\label{identification}
\psi (a) ~&=&~ \psi (b) \nonumber \\
\frac{\partial \psi}{\partial x} (a) ~&=&~ -\frac{\partial \psi}{\partial x} (b)
\end{eqnarray}
Indeed, from the point of view of the wave function Eqn.\,(\ref{identification}) ensures that $a$ and $b$ are in effect the same point (with reversed orientation for the second ray), so that we obtain the quantum mechanics of a line.

A more general framework includes both Eqn.\,(\ref{robin}) and Eqn.\,(\ref{identification}) as special cases, and allows smooth interpolation between then.  To capture the underlying formal structure, we define the vector
\begin{equation}\label{bcVector}
u ~\equiv~ \left(\begin{array}{cccc} \psi(a) & \psi(b) &\frac{\partial \psi}{\partial x} (a)  & \frac{\partial \psi}{\partial x} (b) \end{array}\right)^T
\end{equation}
and matrix
\begin{eqnarray}\label{JMatrix}
J ~&=&~ \left(\begin{array}{cccc}0 & 0 & 1 & 0 \\0 & 0 & 0 & 1  \\ -1 & 0 & 0 & 0 \\0 & -1 & 0 & 0\end{array}\right) .\nonumber
%~&=&~ \left(\begin{array}{cc}0 &1 \\ -1& 0 \end{array}\right)
\end{eqnarray}
With these notations, the current conservation condition reads
\begin{eqnarray}
u^\dagger J u ~&=&~  0  ~~~\Rightarrow \nonumber \\
w^\dagger J v ~&=&~  0
\end{eqnarray}
where in the second line we display the polarized form.
Now we implement boundary conditions by projecting the allowed vectors $v$ to be of the form $v = \Pi \xi$, with a fixed real projection operator $\Pi$ and an arbitrary numerical vector $\xi$.  In this formulation, for example, we represent Dirichlet boundary conditions and  identification boundary conditions respectively with
\begin{eqnarray}
\Pi_{\rm D} ~&=&~ \left(\begin{array}{cccc}0 & 0 & 0 & 0 \\0 & 0 & 0 & 0 \\0 & 0 & 1 & 0 \\0 & 0 & 0 & 1 \end{array}\right) \label{dirichletPi} \\
\Pi_{\rm =} ~&=&~ \mbox{$\frac12$}\left(\begin{array}{cccc}1 & 1 & 0 & 0 \\1  & 1 & 0 & 0 \\ 0 & 0 & 1 & -1 \\0 & 0 & -1 & 1\end{array}\right) \label{identificationPi}
\end{eqnarray}

We will have current conservation within the allowed Hilbert space if
\begin{equation}\label{matrixCC}
\Pi^\dagger J \Pi ~=~ 0
\end{equation}
To obtain a conventional eigenvalue problem we want two conditions, so that $\Pi$ will represent projection onto a two-dimensional subspace.  Projection onto a subspace of larger dimension will not permit one to satisfy Eqn.\,(\ref{matrixCC}).  Projection onto a subspace of lower dimension leads to an unconventional, though not necessarily inconsistent, eigenvalue problem and Hilbert space.  The antisymmetric matrix $J$ implements a hermitean symplectic inner product, and maximal (here, two-dimensional) projections satisfying Eqn.\,(\ref{matrixCC}) define Lagrangian subspaces with respect to $J$.

{\it Interpolation\/}: We can interpolate continuously between identification boundary conditions $\Pi_{=}$ at $\theta =0$ and Dirichlet boundary conditions $\Pi_{D}$ at $\theta = \pm \pi/2$, while satisfying the requirements of Eqn.\,(\ref{matrixCC}) and rank($\Pi$)~=~2 throughout, in the manner
\begin{equation}\label{splitJoinInterpolation}
\Pi (\theta ) ~=~
\mbox{$\frac12$} \left(\begin{array}{cccc } c^2 & c^2 & cs & cs
\\ c^2 & c^2 & cs & cs
\\ cs & cs & 1+ s^2  & - c^2
\\ cs & cs & - c^2 & 1 + s^2  \end{array}\right) \
\end{equation}
(abbreviating $\cos \theta \rightarrow c$,  $\sin \theta \rightarrow s$).
$\Pi (\theta)$ implements projection onto the subspace spanned by $(0, 0, 1, -1)^T$ and $(c, c, s,  s)^T$.
Note that for intermediate values of $\theta$, this subspace does not have a purely  geometric interpretation in physical space.  

This basic joining -- or, read in the opposite sense, splitting -- operation can be used as a module to implement more complex topological transformations, such as the pinching off of a closed ``baby universe'' -- that is, a circle -- as smooth interpolations among consistent, reasonably conventional quantum mechanical systems.    Figure 1 illustrates that geometric possibility.   (A different interpolation between one circle and two appears in \cite{bala}.)
\vspace{0.1in}
\begin{figure}[ht]
\includegraphics[scale=1]{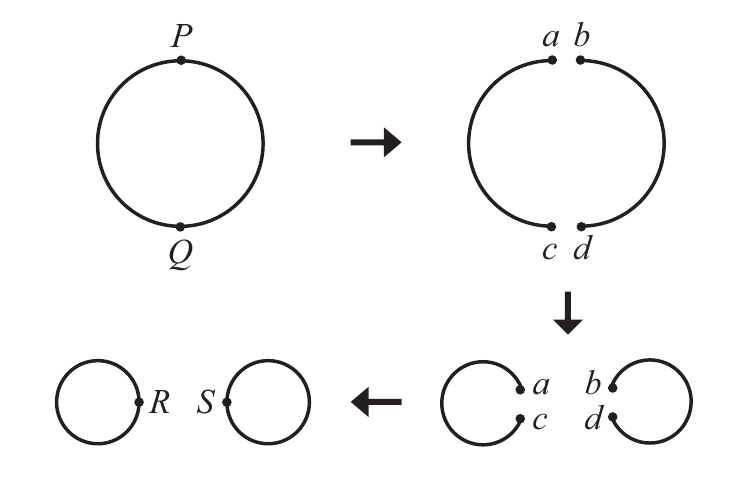}
\caption{Starting with a circle, breaking at $P \rightarrow a, b$ and $Q \rightarrow c,d$, then joining $a,c \rightarrow R$ and $b, d \rightarrow S$ yields two circles.}
\end{figure}

The concept of smooth interpolation between topologies can be extended to higher dimensions.  In that context, we need conditions on the wave function and its normal derivative.   In Figure 2 we illustrate a continuous ``surgery'' leading from two spheres to a torus plus four waste disks.
\begin{figure}[ht]
\includegraphics[scale=.1]{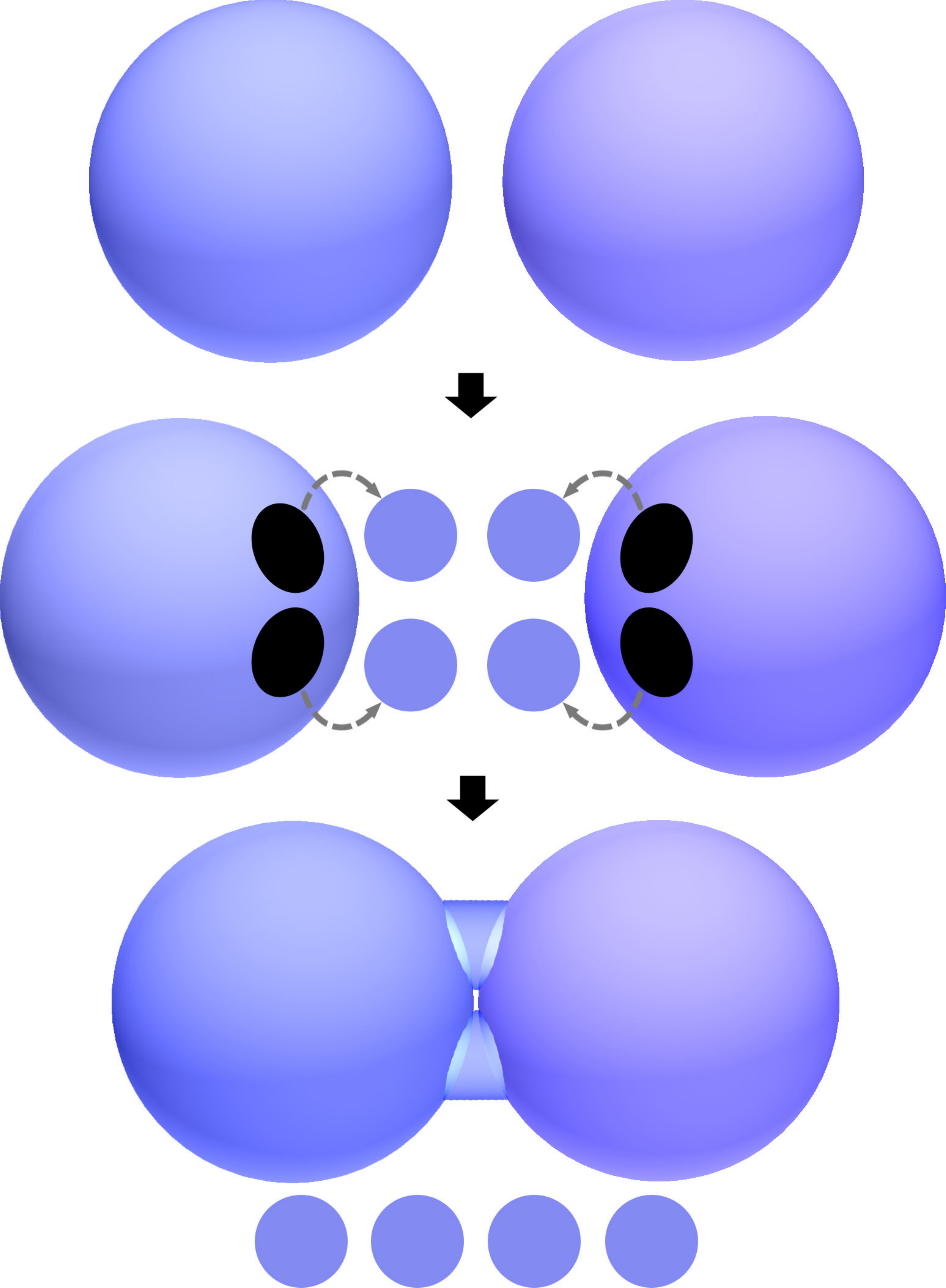}
\caption{By passing from identification to Dirichlet boundary conditions on four small circles, discarding the interior disks, and passing from Dirichlet to identification conditions on two circular boundaries, we interpolate continuously between quantum mechanics on two spheres and quantum mechanics on a torus plus four disks.}
\end{figure}
In these constructions, the interpolation angle $\theta$ can, but need not, be a (spatially) constant function on each bounding circle.

{\it Groups and Geometry\/}:  We now briefly elaborate on the general framework wherein interpolation through boundary conditions is naturally discussed.  (This structure arises, and has been studied, in the context of quantum graphs \cite{schrader}.  Related structures appear in \cite{bala} and \cite{Asorey}.)

Consider a network of wires with junctions of arbitrary multiplicity. 
We want to insure that no net probability flows into a given junction where $n$ segments terminate at points $x_j$.  (Given $p$ wires we have $n=2p$ possible terminals.)  If we  orient all derivatives to define inward normals (that is, oriented away from the termini), then the necessary and sufficient condition for current conservation takes the form
\begin{equation}\label{conservation}
w^\dagger  J v ~=~ 0
\end{equation}
where
\begin{equation}
v^T ~\equiv~ (\psi_1(x_1), \psi_1(x_2), ... , \psi_1 (x_n), \psi_1^\prime (x_1), \psi_1^\prime (x_2), ... , \psi_1^\prime (x_n))
\end{equation}
and similarly for $w$, with $\psi_1 \rightarrow \psi_2$, and
\begin{equation}
 J ~=~ \left(\begin{array}{cc}0 & 1 \\-1 & 0\end{array}\right)
\end{equation}
where each entry denotes an $n\times n$ matrix.
Again $J$ implements a hermitean symplectic inner product.

Now we want to insure Eqn.\,(\ref{conservation}) by making a projection $\Pi$ of the allowed vectors onto an $n$-dimensional subspace.   So we require
\begin{equation}\label{piCondition}
\Pi^\dagger  J \Pi ~=~ 0
\end{equation}

If we have a solution $\Pi$, then we can look for others by varying the subspace, so $\Pi \rightarrow { U} \Pi { U}^{-1}$.  Without loss of generality, we can take $\Pi$ to be hermitean, and $U$ unitary.   Eqn.\,(\ref{piCondition}) goes over into
\begin{equation}\label{uCondition}
{ U}^\dagger  J { U} ~=~  J
\end{equation}

If we write the ${ U} = 1 + \epsilon M$ for infinitesimal $\epsilon$ then Eqn.\,(\ref{uCondition}) gives
\begin{equation}
M ~=~ \left(\begin{array}{cc}A & B_h \\ C_h  & - A^\dagger \end{array}\right)
\end{equation}
where $B_h$ and $C_h$ are hermitean.  Combining this with unitarity ({\it i.e.}, $M$ antihermitean) we find
\begin{equation}\label{mCondition}
M ~=~  \left(\begin{array}{cc}A_a & B_h \\-B_h & A_a \end{array}\right)
\end{equation}
where $A_a$ is anti-hermitean.
This form of $M$ spans two commuting ${\rm U}(n)$ Lie algebras, {\it i.e.}
\begin{equation}
M_{\pm} ~=~ \left(\begin{array}{cc} A_a  & \pm i A_a  \\ \mp i A_a  & A_a \end{array}\right)
\end{equation}
Exponentiating, we see that the ``unitary hermitean symplectic'' group is isomorphic to ${\rm U}(n)\times {\rm U}(n)$, embedded in ${\rm U}(2n)$ as
\begin{equation}\label{cosetForm}
(U_1, U_2) ~\leftrightarrow~ \mbox{$\frac12$}\left(\begin{array}{cc}U_1+ U_2 & i (U_1-U_2) \\-i(U_1-U_2) & U_1 + U_2 \end{array}\right)
\end{equation}

We get the geometry of allowed subspaces as a coset space, by choosing a reference $\Pi_0$ and dividing out by the isotropy subgroup of $ U$ matrices for which ${ U}^\dagger \Pi_0 { U} = \Pi_0$.   Choosing pure Dirichlet
\begin{equation}
\Pi_0 ~=~ \left(\begin{array}{cc}0 & 0 \\ 0 & 1 \end{array}\right)
\end{equation}
as reference, we find that the diagonal subgroup is the isotropy subgroup.   Thus the geometry of the set of Lagrangian subspaces is the geometry of ${\rm U}(n) \times {\rm U}(n)$ divided by its diagonal subgroup, {\it i.e.} the coset space ${\rm U}(n)\times {\rm U}(n)/{\rm U}(n)_\Delta$.   For the projector we have
\begin{eqnarray}
\Pi ~=~\mbox{$\frac12$}
\left(
\begin{array}{cc}
1 - \frac{W+W^\dag}{2} & i \frac{W-W^\dag}{2} \\
i \frac{W-W^\dag}{2} & 1 + \frac{W+W^\dag}{2}
\end{array}
\right)
\end{eqnarray}
where $W \equiv U_1^\dagger U_2$.

If we restrict to boundary conditions that do not mix the wave functions with their derivatives, then we have $B=0$ in Eqn.\,(\ref{mCondition}).  This picks out the diagonal subgroup ${\rm U}(n)_\Delta$ of the general ${\rm U}(n)\times {\rm U}(n)$, with $U_1 = U_2$.  This group is naturally associated with boundary conditions of the form
\begin{equation}\label{identificationBC}
\Pi  ~=~ \left(\begin{array}{cc}R & 0 \\ 0 & S \end{array}\right)
\end{equation}
with $R, S$ projection operators that treat wave functions and derivatives separately.   Returning to the general case, for current conservation we want $RS = 0$, and for a good eigenvalue problem we want ${\rm rank} (R) + {\rm rank} (S) = n$, so as projections $R+S=1$. The diagonal subgroup will preserve the rank of $R$ and $S$ separately, so we have $n+1$ distinct orbits.  Geometrically, the orbits are Grassmannians ${\rm U}(n)/{\rm U}(k)\times {\rm U}(n-k)$.

For example, for $n=2$ and ${\rm rank} (R) =  {\rm rank} (S) = 1$ we can take the reference
\begin{eqnarray}
R_0 ~&=&~ \mbox{$\frac12$}\left(\begin{array}{cc}1 & 1 \\ 1&  1\end{array}\right) \nonumber \\
S_0 ~&=&~ \mbox{$\frac12$} \left(\begin{array}{cc}1 & -1 \\ - 1&  1\end{array}\right)
\end{eqnarray}
which implements the identification boundary conditions of Eqn.~(\ref{identificationPi}).
For this choice of $R_0, S_0$  the isotropy subgroup is generated by $(e^{i\theta}, e^{i \theta})$ and $(e^{i \theta \sigma_1}, e^{i \theta \sigma_1} )$, and the orbit geometry is ${\rm SU}(2)/{\rm U}(1)$.   For $R_0$ the identity and $S_0 = 0$, or {\it vice versa}, corresponding to pure Neumann or pure Dirchlet, respectively, we get an orbit consisting of a single point.

{\it Examples}: For the purpose of studying dynamical topology change, we would like to consider time-dependent boundary conditions as providing a kind of external field.
Below we will indicate how this can be done rigorously; here we provisionally assume adiabatic evolution, and check its internal consistency.   Within the space defined by $\Pi$ the instantaneous Hamiltonian defines a normal eigenvalue problem, and the amplitudes on different levels obey the usual quantum dynamical equations.

Returning to the example of fission of the interval $[ 0, 2 \pi ]$ into two equal intervals, with a free particle Hamiltonian $H= -\bigl( \frac{\partial}{\partial x} \bigr)^2$, we can carry forward explicit calculations  in the adiabatic approximation.   It is convenient to use coordinates $0 \leq x \leq 2\pi$ for the mother interval and $0 \leq y =x \leq \pi$, $0 \leq z = 2\pi -x \le \pi$ on the daughters.   We maintain Dirichlet boundary conditions at $x= 0 , 2\pi$, $y= 0$, $z=0$ throughout, and evolve from identification to Dirichlet at $x= y = z = \pi$, following our earlier prescription.   Eigenfunctions can be taken real, and then follow the forms $(\sin ky , \mp \sin kz)$.  We have the boundary condition
\begin{eqnarray}
&{}& (\psi (y=\pi), \psi (z=\pi), \psi^\prime (y=\pi) ,  \psi^\prime (z=\pi) )  \nonumber \\
&&~~\propto~ (a \cos \theta, a \cos \theta,  a \sin \theta + b ,  a \sin \theta - b )
\end{eqnarray}

For the upper sign $(\sin ky, - \sin kz)$ the solutions are $k = l$ where $l$ is a positive integer, independent of $\theta$, with $a=0$.  In terms of $x$ these modes are simply $\sin kx$.  They are antisymmetric around $x= \pi$, vanish there and are smooth.

For the lower sign $(\sin ky, \sin kz)$ the solutions have $b=0$ and $k \cot \, k\pi \, = \, \tan \, \theta$.  In terms of $x$ these modes take the form $\theta (\pi -x) \sin \, kx  + \theta (x - \pi) \sin \, k (2 \pi -x) $.  They are symmetric around $x = \pi$, and have kinks there.   At $\theta = 0$ we have $k = l -1/2$, where $l$ is a positive integer, and $k$ evolves continuously until $k=l$ at $\theta = - \pi/2$.

Since the gap between states of the same symmetry never closes, adiabatic evolution of states is well defined, and will govern sufficiently slow evolution in the regulated dynamical framework described below.    There is no geometric (Berry) phase, since we can use bases of real eigenfunctions throughout.

The adiabatic evolution of the eigenfunctions has a simple intuitive interpretation in terms of node adjustment.  Eigenfunctions of the first class (upper sign) already have a node at the midpoint, so they are pre-adapted to Dirichlet boundary conditions, and do not change.  Eigenfunctions of the second class (lower sign) have a local maximum or minimum at the midpoint.  As they evolve adiabatically, that extremal value is pushed down toward zero.  The interior nodes move slightly in the direction away from the midpoint, but do not disappear, and no additional nodes are created.

Interesting complications occur if one considers fission away from the midpoint.
Node induction implies that wave functions of both kinds will be adjusted.
If the point of fission only deviates slightly from the midpoint, then among the low energy states, the modes with $k=l$ are pushed into the narrower interval and their energy floats up slightly, whereas the modes with $k=l-\frac 12$ are pushed into the wider interval, ending up with $k$ that are smaller but close to $l$. The energy gap narrows as we approach the Dirichlet point, which requires slower evolution. In the other extreme, if we split at very small $x = \epsilon$, the low energy wave functions are easily adaptable: they simply move out of the interval $[0,\epsilon]$.

{\it Quantum Entropy and Entanglement\/}:  Our midpoint fission of the interval yields a mapping of Hilbert spaces $L^2([0,2\pi]) \rightarrow L^2([0,\pi])\oplus L^2([0,\pi])$.  (Note: The direct sum appears here, {\it not\/} the tensor product.)   In that notation, we have an informative basis
\begin{eqnarray}\label{uvModes}
u_p ~&\equiv&~ \frac{1}{\sqrt{2\pi}}  ( \sin px + \sin \frac{2p-1}{2} x ) ~ \rightarrow ~ \sqrt{\frac{2}{\pi}} \sin py \oplus 0 \nonumber \\
v_p ~&\equiv&~ \frac{1}{\sqrt{2\pi}} (\sin px - \sin \frac{2p-1}{2} x)  ~\rightarrow ~0 \oplus  \sqrt{\frac{2}{\pi}} \sin pz \nonumber \\
&{}&
\end{eqnarray}
with $p$ running through the positive integers.  (Note that since $u_p, v_p$ are not eigenfunctions until the fission is complete, their coefficients in a dynamical wave function may have complicated time dependence prior to then.  The mapping of Eqn.\,(\ref{uvModes}) is defined structurally, rather than directly dynamically.)

Now if we consider an observer confined to the first daughter interval $L^2 ([0, \pi])$, since all the possible operations are also confined to it, the appropriate density matrix would have two separate sectors, corresponding to one-particle states $\sin py$ and the no-particle state. We can simply parameterize them with indices $p,\ 0$. Starting with the wave function $\sum_p \alpha_p u_p + \beta_p v_p$ we find the density matrix
\begin{eqnarray}
\rho_{(q,p)} ~&=&~ \alpha_q^* \alpha_p \nonumber \\
\rho_{(0,0)} ~&=&~ \sum_p |\beta_p |^2 \nonumber \\
\rho_{(q, 0)} ~&=&~ \rho_{(0, p)} ~=~ 0 \label{confinedDensityMatrix}
\end{eqnarray}
This is not a pure-state density matrix, but it is close to one: a renormalized version of $\rho_{(p,q)}$, dropping the no-particle entries, would be a pure-state matrix. A nonzero quantum entropy is associated with it
\begin{equation}
- {\rm Tr}( \rho \log_2 \rho ) ~=~ -r \log_2 r -  (1-r) \log_2 (1-r)
\end{equation}
where $r \equiv \sum_p | \beta_p |^2$.
It is maximized at one bit when $\sum_p | \beta_p |^2 = \frac{1}{2}$.
This result makes intuitive sense, because von Neumann entropy is the intrinsic part of the uncertainty of a quantum state (as opposed the excess classical entropy unveiled by measurement), and confinement to one subspace simply deprives the observer of the control over the other intrinsically. More formally, implicitly underlying the knowledge of particle number is the fact that the observer has applied a measurement by $P_1,\ P_2$, orthogonal projectors onto the two subintervals respectively, turning the original $\rho$ to $P_1 \rho P_1 + P_2 \rho P_2$. Such measurements always increase entropy (if not leaving it unchanged).

Already in the two-particle case we find much richer structure.   The mother Hilbert space is $L^2([0, 2\pi]) \otimes L^2([0, 2\pi])$, which fissions into $\bigl( L^2([0,\pi])\oplus L^2([0,\pi]) \bigr) \otimes \bigl( L^2([0,\pi])\oplus L^2([0,\pi]) \bigr)$.   There are four sectors in the product, corresponding to both particles in the first interval, the first particle in the first interval and the second in the second, and so forth.   Using the $u_p, v_p$ basis we can write the initial state as
\begin{equation}
\psi ~=~ \sum_{r,s} ( \alpha_{rs} u_r \otimes u_s + \beta_{rs} u_r \otimes v_s + \gamma_{rs} v_r \otimes u_s + \delta_{rs} v_r \otimes v_s )
\end{equation}
When we form the density matrix for the first interval there will be four sectors, corresponding to both particles in the interval, the first particle only, the second particle only, or no particle.  Their coefficients will be quadratic in $\alpha, \beta, \gamma, \delta$ respectively.   The most interesting cases are when we have one particle, let's say the first.   In an evident notation, we have for that sector the density submatrix
\begin{equation}
\rho_{(q, p)} ~=~ \sum\limits_l \beta_{ql}^* \beta_{pl}
\end{equation}
Since several values of $l$ can make nontrivial contributions, this will generally not be a (renormalized) pure-state density matrix, and we have true entanglement entropy.

In this discussion we have assumed the two particles are distinguishable.  Bosons or fermions can be handled without serious difficulty, by imposing symmetry or antisymmetry on $\alpha, \beta, \gamma, \delta$.

{\it Boundary Conditions As Limiting Interactions}: Boundary conditions can be implemented formally as singular interactions.  For example $\psi (a) = 0$ is enforced, for finite energy states, by the interaction $\Delta H = v | \psi (a) |^2$ as $v \rightarrow \infty$, corresponding to the potential $V(x) = v \delta(x-a)$.  The boundary condition $\psi(a) = \psi(b)$ can be implemented through the energy term $\Delta H = v |\psi (a) - \psi(b)|^2$, which corresponds to a nonlocal potential in the form
\begin{eqnarray}\label{bcInteraction}
&{}&\langle \psi | \Delta H | \psi \rangle ~=~  \\
&{}&v \int dy \int dx \ \psi^*(y) \bigl( \delta(x -a) \delta(y-a)  + \delta(x-b) \delta (y-b)\nonumber \\
&{}& -  \delta(x -a) \delta(y-b) - \delta(x-b) \delta (y-a) \bigr) \psi(x) \nonumber
\end{eqnarray}
or as an operator
\begin{eqnarray}\label{bcOperator}
&{}&(\Delta {\hat H} \psi ) (x) ~=~ \nonumber \\
&{}& v \bigl(  \delta(x-a) \psi (a) + \delta (x-b) \psi(b) \nonumber \\
&{}&- \delta(x-a) \psi (b) - \delta(x-b) \psi (a) \bigr)
\end{eqnarray}
Boundary conditions involving spatial derivatives bring in potentials with derivatives of delta functions.

\vspace{0.1in}
\begin{figure}[ht]
\includegraphics[scale=1]{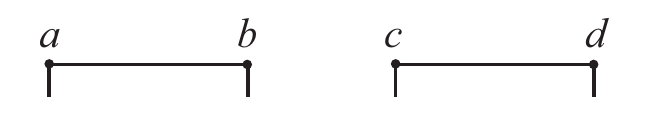}
\caption{By allowing stubs, we can relax the boundary conditions at $a, b, c, d$ while maintaining strict unitarity.  The stubs need not be short, nor separate.}
\end{figure}

For many reasons, including possible practical realization, it is interesting to consider relaxing the conditions on these singular interactions. Specifically, we would rather not insist on strict delta functions or infinite coefficients.   This is not unproblematic, however, since such compromises threaten unitarity.   It is also desirable to have all the dynamics occur within a single master Hilbert space, rather than a time-dependent sequence of different function spaces.  At some cost in elegance we can implement our constructions approximately, while maintaining unitarity exactly, in a single Hilbert space as follows.   Instead of using real terminal points, where we need strict boundary conditions, in our constructions, we let $a, b, c, d, ... $ be interior points, each attached to a stub (See Figure 3).  Instead of using $\delta$ functions (or derivatives of $\delta$ functions) localized near these points in Eqns.\,(\ref{bcInteraction}, \ref{bcOperator}) and their generalizations, we use regular approximations to those functions, {\it e. g}. Gaussians and their derivatives, with large though finite coefficients.   Then wave functions that are supported in the region between the stubs will realize, approximately, the dynamics of topology change we proposed above.   Since the boundary conditions are only enforced approximately some probability will leak into the stubs, but if the approximation to the singular interactions is accurate that leakage will be small.

{\it Application to Quantization}: In a recent paper \cite{branched}, two of us proposed a method for quantizing Lagrangians whose kinetic terms are of the form $L_{\rm kin.} = \frac{1}{4} {\dot x}^4 - \frac{\kappa}{2} {\dot x}^2$.   Standard quantization procedures do not apply (for $\kappa > 0$), since the associated Hamiltonian is singular, with cusps at $p_\pm = \pm \sqrt{\frac{\kappa}{3}}$ linking three smooth branches where $-\infty < p < p_+$, $p _- < p < p_+$,  $p_- <  p < \infty$. (In the interval between the cusps the Hamiltonian, as a function of $p$, is three-valued.)  After introducing the potential $V(x) = \frac{\alpha}{2} x^2$ and making the duality transformation $x \rightarrow p$, $p \rightarrow -x$, the system maps onto the description of a particle with conventional kinetic energy on a singular space.   Introducing the variable
\begin{eqnarray}
\chi ~&\equiv&~ \ \ \ \, x - x_+ +  x_-\ \ \ {\rm for} \ \ \chi \leqslant x_- \nonumber \\
\chi ~&\equiv&~ -x + x_+ + x_- \ \ \  {\rm for} \ \ x_- \leqslant \chi \leqslant x_+ \nonumber \\
\chi ~&\equiv&~ \ \ \ \, x + x_+ - x_- \ \ \ {\rm for} \ \ x_+ \leqslant \chi
\end{eqnarray}
allows us to unfold the space, so now the branches occupy $-\infty < \chi < x_-$, \,$x_- < \chi < x_+$, \,$x_+ < \chi < \infty$, and the salient issue is what boundary conditions to impose.   In \cite{branched}, guided by unitarity and simplicity, we chose what we've here called identification boundary conditions, that is continuity of the wave function and its first derivative at $x_\pm$.   This leads to a consistent quantization of the original classical Lagrangian, wherein motion through the cusps, taking the particle from branch to branch, is unimpeded.

On the other hand classical mechanics for systems of this kind suggests the interest of allowing ``hard wall'' reversals of $x$ at the cusps \cite{classicalTXTal}, which in turn might suggest imposing $\psi(x_- - \epsilon) = \psi(x_- + \epsilon) = \psi(x_+ - \epsilon) = \psi(x_+ + \epsilon)$ at our four endpoints.  These three conditions together with $\psi^\prime(x_- - \epsilon) -  \psi^\prime (x_- + \epsilon) + \psi^\prime (x_+ - \epsilon) - \psi^\prime (x_+ + \epsilon) = 0$ yield a consistent, unitary theory (``ring quantization'').   Another possibility, perhaps suggested by quantization using noncanonical variables \cite{noncanonical}, \cite{branched}, is to impose Dirichlet boundary conditions $\psi(x_- - \epsilon) = \psi(x_- + \epsilon) = \psi(x_+ - \epsilon) = \psi(x_+ + \epsilon) = 0$.    Again, this is consistent with unitarity and realizes the classical correspondence principle away from the cusps.   An alternative interpretation of the hard wall is that a particle starting on the branch between the cusps remains trapped there forever; this suggests the baby universe boundary conditions $\{ \psi(x_- + \epsilon) = \psi(x_+ - \epsilon);\, \psi^\prime(x_- + \epsilon) = \psi^\prime(x_+ - \epsilon)\}$, supplemented with Dirichlet $\psi(x_- - \epsilon)= \psi(x_+ + \epsilon) = 0$ or identification $\{ \psi(x_- - \epsilon) =  \psi(x_+ + \epsilon);\,  \psi^\prime(x_- - \epsilon) =  \psi^\prime(x_+ + \epsilon) \}$ conditions at the ends of the outer branches.

These several choices -- and others -- all fit naturally within the general framework proposed above.  They correspond to different choices of Lagrangian subspaces within the eight-dimensional space spanned by the values of the wave function and its derivative at the four end-points, with respect to the symplectic structure imposed by unitarity.    Each can be reached from any of the others by interpolation through consistent quantum theories.  The circumstance that a single classical theory can give rise to many possible quantum theories arises in QCD, where one has the $\theta$ parameter \cite{theta}; or, more simply, in the quantum mechanics of a particle on a circle, where one has the possibility of fractional angular momentum \cite{fractionalL}.

\bigskip

{\it Comments}:
\begin{enumerate}
\item The central formal ideas of the preceding constructions, including association of quantum theories with Lagrangian subspaces, smooth interpolation between theories based on topologically distinct manifolds by motion in the consequent moduli spaces, and the realization of all these structures approximately through simple quadratic interactions, carry over to interacting many-body systems in a straightforward way.   With some modification, they also apply to other differential operators, {\it e.g}. Dirac, Majorana, Yang-Mills, or Einstein (curvature).   It would be interesting to elucidate the relation between topology and the spectral flow of such operators, including specifically index theory, from this perspective \cite{spectralFlow}.   In a similar spirit, one could calculate charge flows \cite{gw}.
\item Similar structures arise in quantum field theory by a slightly different route, as follows.   In that context we consider wave functionals $\Psi(\phi(x))$, where now the functions $\phi(x)$ live on various manifolds with boundary.    The Hamiltonian typically contains a kinetic piece which is formally
\begin{equation}
H_{\rm grad} ~\propto~ \int ( \nabla \phi )^2
\end{equation}
operating on $\Psi$.  To define an appropriate Hilbert space of functions $\phi$ for which $H_{\rm grad}$ is Hermitean by local (or nearly local) conditions on the boundary, we are led to impose boundary conditions of the kind discussed above.
\item If we allow spatial variations in the surgery constructions then we have, in view of our discussion in ``groups and geometry'', the structure of a ${\rm U}(2)\times {\rm U}(2)/{\rm U}(2)_\Delta$ gauged sigma model on the boundaries.  Such sigma models and their generalizations will arise in compactifications of extra dimensions.  With four noncompact dimensions, instanton-like configurations based on $\pi_3({\rm U}(n))$ can mediate changes in the topology of the compactification manifold.

\item The idea that quantum wave functions potentially describe many worlds gets realized in a precise and tangible form, as the quantum dynamics of (say) a single interval evolves into that of two separate intervals.   Let us begin with a pure state on the mother interval.  After the interval has fissioned, we still have a pure state on the two-interval multiverse.   But observers (or observables) confined to one interval will find it appropriate to use a mixed state, described by a density matrix that traces over dynamical variables localized on the other interval.   An important point is that the {\it dynamics\/} of these mixed states will appear perfectly local and ordinary. In that sense the evolution has revealed that two distinct daughter worlds were implicit in the wave function of one mother.   Note that the quantum entropy induced by the fission depends on the initial state.
\end{enumerate}

%%%%%%%%%%

%%%%%%%

{\it Acknowledgements}:  ADS is supported in part by NSF Grants  PHY-1214341 and PHY-0855614.  FW is supported in part by DOE grant DE-FG02-05ER41360.


\begin{thebibliography}{99}

\bibitem{bala}
  A.~P.~Balachandran, G.~Bimonte, G.~Marmo and A.~Simoni,
  ``Topology change and quantum physics,''
  Nucl.\ Phys.\ B {\bf 446}, 299 (1995)
  [gr-qc/9503046].

  \bibitem{Asorey}
  M.~Asorey, A.~Ibort and G.~Marmo,
  ``Global theory of quantum boundary conditions and topology change,''
  Int.\ J.\ Mod.\ Phys.\ A {\bf 20}, 1001 (2005)
  [hep-th/0403048].

\bibitem{vonNeumann}
J. von Neumann, ``Allgemeine Eigenwerttheorie Hermitescher Funktionaloperatoren," Math. Ann., {\bf 102} (1929) 49.

 \bibitem{ReedSimon}
 M. Reed and B. Simon, {\it Methods of Modern Mathematical Physics, Vol. 2: Fourier Analysis,
Self Adjointness} [Academic Press, 1975].

 \bibitem{schrader} V. Kostrykin and R. Schrader, ``Kirchhoff's rule for quantum wires",
 J. Phys. A: Math. Gen. {\bf 32}, 595.


\bibitem{branched}
  A.~Shapere and F.~Wilczek,
 ``Branched Quantization,''
  arXiv:1207.2677 [quant-ph].
  %%CITATION = ARXIV:1207.2677;%%

\bibitem{classicalTXTal}
  A.~Shapere and F.~Wilczek,
  ``Classical Time Crystals,''
  arXiv:1202.2537 [cond-mat.other].
  %%CITATION = ARXIV:1202.2537;%%

\bibitem{noncanonical}
  L.~Zhao, P.~Yu and W.~Xu,
 ``Hamiltonian description of singular Lagrangian systems with spontaneously broken time translation symmetry,''
  arXiv:1206.2983 [hep-th].
  %%CITATION = ARXIV:1206.2983;%%

  \bibitem{Ohya}
  S.~Ohya,
 ``Path Integral Junctions,''
  J.\ Phys.\ A A {\bf 45}, 255305 (2012)
  [arXiv:1201.5115 [hep-th]].

 \bibitem{theta}
 C.~Callan, R.~Dashen and D.~Gross ``Structure of the Gauge Theory Vacuum'' Phys. Lett. {\bf B63} 334 (1976);
R. Jackiw and C. Rebbi,  ``Vacuum Periodicity in a Yang-Mills Quantum Theory",
Phys.~Rev.~Lett. {\bf 37} (1976) 172.

 \bibitem{fractionalL}
F.~Wilczek ``Magnetic Flux, Angular Momentum, and Statistics'' Phys. Rev. Lett. {\bf 48} 1144 (1982).

\bibitem{spectralFlow}
M.~ Atiyah, V.~Patodi and I.~Singer, ``Spectral asymmetry and Riemannian
geometry'', Math. Proc. Camb. Phil. Soc. {\bf 79} 99 (1976).

\bibitem{gw}
J. Goldstone and F. Wilczek,
``Fractional Quantum Numbers on Solitons",
Phys.~Rev.~Lett. {\bf 47}  (1981) 986.

\end{thebibliography}
\end{document}